\documentclass[a4paper]{spie}  

\usepackage{amsmath,amsfonts,amssymb}
\usepackage[colorlinks=true, allcolors=blue]{hyperref}
\usepackage{booktabs} 
\title{Enhancing User Intent for Recommendation Systems via Large
Language Models}

\author[a*]{Xiaochuan Xu}
\author[a]{Zeqiu Xu}
\author[a]{Peiyang Yu}
\author[b]{Jiani Wang}
\affil[a]{Information Networking Institute, Carnegie Mellon University, 5000 Forbes Avenue, Pittsburgh, PA 15213, USA; \textsuperscript{b}Department of Computer Science, Stanford University, 450 Jane Stanford Way, Stanford, CA 94305, USA}
\authorinfo{* xiaochux@alumni.cmu.edu}

\begin{document} 
\maketitle

\begin{abstract}
Recommendation systems play a critical role in enhancing user experience and engagement in various online platforms. Traditional methods, such as Collaborative Filtering (CF) and Content-Based Filtering (CBF), rely heavily on past user interactions or item features. However, these models often fail to capture the dynamic and evolving nature of user preferences. To address these limitations, we propose DUIP (Dynamic User Intent Prediction), a novel framework that combines LSTM networks with Large Language Models (LLMs) to dynamically capture user intent and generate personalized item recommendations. The LSTM component models the sequential and temporal dependencies of user behavior, while the LLM utilizes the LSTM-generated prompts to predict the next item of interest. Experimental results on three diverse datasets—ML-1M, Games, and Bundle—show that DUIP outperforms a wide range of baseline models, demonstrating its ability to handle the cold-start problem and real-time intent adaptation. The integration of dynamic prompts based on recent user interactions allows DUIP to provide more accurate, context-aware, and personalized recommendations. Our findings suggest that DUIP is a promising approach for next-generation recommendation systems, with potential for further improvements in cross-modal recommendations and scalability. 
\end{abstract}


\section{INTRODUCTION}
\label{sec:intro}  

Recommendation systems are critical in various online services, from e-commerce and social media platforms to content streaming services. These systems aim to predict and suggest items that align with user preferences, improving user experience and engagement. Traditional recommendation algorithms typically rely on Collaborative Filtering (CF) and Content-Based Filtering (CBF) methods. CF uses historical interaction data, such as user-item interactions, to predict user preferences, while CBF relies on item features to match users with relevant content \cite{BONDEVIK2024122166}\cite{wang2024trustworthy}. Despite their widespread adoption, these approaches often fall short in capturing the dynamic and context-dependent nature of user preferences, especially in real-time context.

In recent years, there has been significant interest in improving recommendation systems by incorporating more sophisticated models capable of understanding user intent and context. Traditional methods often assume that user preferences are static, based on past behavior, but emerging techniques have begun leveraging contextual information, such as user sentiment\cite{yu2021text}, emotional state, event information extraction\cite{liu2024beyond,liu2023enhancing} or temporal factors, to dynamically adjust recommendations. Large Language Models (LLMs) like GPT and BERT have shown exceptional promise in understanding complex user input and context\cite{yu2024applications}, enabling them to extract deeper insights from user-generated content such as reviews, comments, and queries \cite{zhu2024collaborative}. Recent studies have also explored the integration of LLMs with recommendation systems to improve personalized recommendations by leveraging natural language understanding \cite{ren2024representation}.

Despite the progress, several challenges persist in the development of next-generation recommendation systems. One major issue is that traditional models often fail to accurately capture dynamic user intent, which can change over time and across contexts. While context-aware recommendation systems have made strides in integrating real-time factors like sentiment or social context, they still lack a robust mechanism for predicting and adapting to user intent shifts in a timely and reliable manner \cite{bakhshizadeh2024supporting}.

Furthermore, the cold-start problem remains a significant hurdle, especially for new users or items where limited interaction data is available, making it difficult to infer user preferences effectively \cite{chaimalas2023bootstrapped}. There is a need for systems that can dynamically infer user intent from sparse data, adapt recommendations in real-time, and seamlessly integrate cross-modal information (e.g., text, images, social context) \cite{li2023multi}.

To address these challenges, this paper proposes a novel framework that integrates LSTM networks and LLMs for dynamic user intent prediction, named \textbf{DUIP}. In our approach, an LSTM model is first employed to capture the dynamic intent of the user based on their recent interactions. The output from the LSTM, which serves as a prompt template, is then fed into a Large Language Model (such as GPT-2) to predict the next item the user may be interested in. By using this approach, we combine the LSTM's ability to capture user intent over time with the LLM’s powerful language understanding to make more contextually aware and accurate predictions. This model not only addresses the cold-start problem by using dynamic prompts but also enhances personalization by continuously adapting to shifting user needs and preferences.

The rest of the paper is organized as follows: Section 2 provides an overview of the related work in the area of recommendation systems, focusing on approaches for dynamic intent modeling, context-aware recommendations, and the application of Large Language Models in these domains. Section 3 presents the detailed methodology of the proposed approach, explaining the integration of LSTM networks with LLMs for dynamic user intent prediction and item recommendation. Section 4 describes the experimental setup, including datasets, evaluation metrics, and comparison with baseline methods. Finally, Section 5 concludes the paper by summarizing the findings, discussing the implications of the results, and suggesting directions for future research.

\section{RELATED WORK}

\subsection{Intent Modeling in Recommendation Systems}

Intent modeling has become a crucial research direction in recommendation systems, aiming to better understand and predict user needs. Traditional recommendation methods, such as Collaborative Filtering and Content-Based Filtering, rely heavily on user historical data \cite{BONDEVIK2024122166}. However, these methods often fail to capture real-time user needs and emotional fluctuations, which has prompted interest in dynamic intent modeling approaches \cite{chen2022intent}. Sentiment analysis has been widely used to understand user intent, as it helps identify users' emotional states and adjusts recommendations accordingly \cite{sun2024large}. Several studies have focused on integrating contextual information such as emotional states, location, and social context to enhance recommendation systems \cite{ren2024representation}.

\subsection{Application of Large Language Models in Recommendation Systems}

The application of Large Language Models (LLMs), such as GPT and BERT, in recommendation systems has been gaining attention due to their ability to process and understand large amounts of text data. These models can generate more accurate and personalized recommendations by understanding the context and intent behind user-generated content \cite{ren2024representation}. The main research directions in this area include,

Sentiment-Aware Recommendation: By using LLMs to perform sentiment analysis on user reviews and social media data, systems can infer user emotional states and adjust recommendations accordingly \cite{bakhshizadeh2024supporting}.

Conversational Recommendation Systems: LLMs are applied in dialog-based recommendation systems, allowing the system to generate personalized recommendations based on real-time feedback from users \cite{sun2024large}.

Generative Recommendation: Some studies explore using LLMs for generative recommendation, where the model generates personalized recommendations directly from user inputs such as search queries or reviews, instead of matching
items from historical data \cite{zhu2024collaborative}.

However, most existing research is primarily focused on sentiment analysis and conversational recommendation, with little emphasis on how LLMs can be used to enhance real-time user intent modeling and dynamically adjust recommendation strategies.

\subsection{Intent Inference and Context-Aware Recommendations}
Intent inference and context-aware recommendation have become core topics in modern recommendation system research. Unlike traditional static methods, dynamic intent inference captures real-time user needs and provides personalized recommendations accordingly \cite{wang2024trustworthy}. Incorporating multi-modal information (e.g., text, images, social
network data) allows recommendation systems to better understand users' current intent and emotional state \cite{ren2024representation}.

Context-Aware Systems: Context-aware recommendation systems integrate user sentiment, social context, and emotional states to provide more accurate and personalized recommendations \cite{bakhshizadeh2024supporting}. For example, combining sentiment analysis with historical data can improve recommendation responsiveness and user satisfaction.

Cross-Modal Recommendation Systems: There has been growing interest in cross-modal recommendation, which combines text, image, and video data to provide more diverse and personalized recommendations \cite{BONDEVIK2024122166}.

\section{METHODOLOGY}

This section introduces a novel methodology combining Long Short-Term Memory (LSTM) networks and Large Language Models (LLMs) to model dynamic user intent and generate personalized item recommendations. The key idea is to leverage the LSTM's hidden state to create a learnable soft prompt template that will guide the LLM in predicting
the next item the user is likely to interact with, based on the user's evolving preferences.

\subsection{Dynamic User Intent Modeling Using LSTM}
The primary objective of this step is to capture dynamic user intent—the preferences and interests that change over time—as users interact with the system. To achieve this, we utilize LSTM networks, which excel at modeling sequential data and capturing 
temporal dependencies in user behavior.

Consider the sequence of user interactions as a time-series of events:
\begin{equation}
	X=\left\{x_{1}, x_{2}, \ldots, x_{t}\right\}
\end{equation}

where each $x_t$ represents the feature vector associated with an item interaction at time t. The feature vector $x_t$ contains various pieces of information, such as: item-specific features (e.g., item ID, category, price) and user context (e.g., time of interaction, session length, device used).
At each timestep, the LSTM network processes this sequence of interactions and produces a hidden state $h_t$, which encapsulates the user's evolving intent:
\begin{equation}\renewcommand\arraystretch{1.5} 
	\begin{array}{c}
	\mathbf{I}_{t}=\sigma\left(\mathbf{X}_{t} \mathbf{W}_{x i}+\mathbf{H}_{t-1} \mathbf{W}_{h i}+\mathbf{b}_{i}\right), \\
	\mathbf{F}_{t}=\sigma\left(\mathbf{X}_{t} \mathbf{W}_{x f}+\mathbf{H}_{t-1} \mathbf{W}_{h f}+\mathbf{b}_{f}\right), \\
	\mathbf{0}_{t}=\sigma\left(\mathbf{X}_{t} \mathbf{W}_{x o}+\mathbf{H}_{t-1} \mathbf{W}_{h o}+\mathbf{b}_{o}\right), \\
	\tilde{\mathbf{C}}_{t}=\tanh \left(\mathbf{X}_{t} \mathbf{W}_{x c}+\mathbf{H}_{t-1} \mathbf{W}_{h c}+\mathbf{b}_{c}\right) \\
	C_{t}=\mathbf{F}_{t} \odot C_{t-1}+\mathbf{I}_{t} \odot \tilde{\mathbf{C}}_{t} \\
	H_{t}=\mathbf{O}_{t} \odot \tanh C_{t}
\end{array}
\end{equation}
where $\mathbf{X}_t$ is input, $\mathbf{I}_t$, $\mathbf{F}_t$, $\mathbf{O}_t$, ${\tilde{\mathbf{C}}}_t$ and $C_t$ are input gate, forgetting gate, output gate, candidate memory cell and memory cell respectively. $\mathbf{W}_{\ast i}$ are weight parameter and $\mathbf{b}_\ast$ are biases. $\sigma $ and tanh is activation function.

The hidden state $h_t$ encodes both short-term preferences (e.g., recently viewed items) and long-term patterns (e.g., sustained interest in certain categories), allowing the model to adapt to changing user behaviors.

\subsection{Soft Prompt Construction with LSTM’s Hidden State}
Once the LSTM has generated the hidden state $h_t$, the next step is to transform this hidden state into a soft prompt that will guide the LLM in making predictions. The soft prompt is a learnable representation that adapts to the user's evolving preferences and provides contextual information for the LLM to use when predicting the next item. 

To form the soft prompt $P$, we first utilize the LSTM's hidden state $h_t $, which encodes the user's current intent. The hidden state is passed through a learnable transformation function $f(\cdot) $, which converts it into a structured vector suitable for input to the LLM.
\begin{equation}
	P=f\left(h_{t}\right)
\end{equation}
where $P$ is the soft prompt derived from the LSTM's hidden state $h_t$, and $f(\cdot)$ is a transformation function (such as a linear layer or multi-layer perceptron). This function maps the LSTM hidden state to a more interpretable format that encapsulates the user's preferences and intent in a contextual vector.

In addition to the soft prompt, we also include hard prompts, which are fixed or predefined information about the user or system. Hard prompts can include static information such as:

\begin{itemize}
	\item User interaction history: previous items the user interacted with.
	\item Item metadata: item categories, tags, etc.
\end{itemize}

The final soft prompt $P$ is formed by combining both the dynamic (soft) prompt from the LSTM and the static (hard) prompts representing user-item interactions:
\begin{equation}
	\begin{array}{r}
		P=\left\langle\text { user }_{i}\right\rangle \text { has interacted with }\{z\} \text { soft } \\
		+ \text { hard prompt }\left\langle r, p_{i}\right\rangle
	\end{array}
\end{equation}
where $\left\langle\text { user }_{i}\right\rangle$ represents the user identity, $\left\{z\right\}$ represents the dynamic soft prompt constructed from the LSTM's hidden state $h_t$, and $\left\langle r, p_{i}\right\rangle$ represents hard prompts that encode historical information about the user's interactions with specific items.

By combining the soft and hard prompts, the model is able to adapt dynamically to the user's current preferences (via the LSTM) while also using historical context to inform the predictions.

Consider the case where the user has recently interacted with laptops and accessories. The LSTM's hidden state $h_t$ could encode this intent, and the transformation function $f\left(h_t\right)$ would produce a soft prompt that reflects the user's interest in laptops. For example, the generated soft prompt might look like:
\begin{equation}
	\begin{array}{c}
		P=\left\langle\text {user}_{i}\right\rangle \text { has interacted with \{laptop, accessories\}} \\ \text{ soft } 
		+ \text {hard prompt}\left\langle\text {item}_{j}, \text {item}_{k}\right\rangle
	\end{array}
\end{equation}

This prompt now includes both the dynamic intent from the LSTM and historical interaction data, which can be passed to the LLM.
\subsection{LLM for Item Prediction}
Once the soft prompt P has been constructed, it is input into the Large Language Model (LLM), such as GPT-2, to predict the next item the user might be interested in. The LLM uses the contextual information from the prompt to generate predictions.

The LLM processes the soft prompt P and generates a ranked list of predicted items that are relevant to the user's current intent. By conditioning the LLM on the dynamic soft prompt, the model can generate context-aware recommendations.

For example, if the prompt indicates that the user is interested in laptops and accessories, the LLM might generate recommendations like:
\begin{equation}
	\begin{aligned}
		&	\text{Recommendations} \\
		= &  \{ ``\text{Laptop Stand}",``\text{Wireless Mouse}",``\text{Laptop Bag}" \}
	\end{aligned}
\end{equation}

The LLM generates the next item $\hat{y}$ by selecting the item with the highest conditional probability:
\begin{equation}
	\hat{y}=\arg \max _{y \in Y} P(y \mid P)
\end{equation}
where $Y$ is the set of candidate items, $P\left(y\mid P\right)$ is the probability of item y being the next relevant item, conditioned on the prompt P This probabilistic framework ensures that the LLM can select the most contextually relevant item, guided by the user's current preferences and historical interaction data.

In this methodology, we integrate LSTM-based dynamic user intent modeling with LLM-based item prediction. The LSTM captures the user's evolving preferences over time, while the soft prompt generated from the LSTM's hidden state guides the LLM in predicting the next relevant item. By combining both soft and hard prompts, our system ensures that the recommendations are not only personalized but also contextually relevant based on the user's interaction history and current intent.
\section{EXPERIMENTS}
In our experiments, we use three real-world datasets from diverse domains to evaluate the performance of our
recommendation framework.

\subsection{Datasets}
The MovieLens-1M (ML-1M) dataset contains 3,416 items and 784,860 sessions, with an average session length of 6.85 and a density indicator of 1573.86. This dataset consists of user ratings for movies and is commonly used for evaluating collaborative filtering approaches.

The Amazon Games dataset is a subcategory of the larger Amazon dataset, containing 17,389 items and 100,018 sessions. The average session length is 4.18, and the density indicator is 24.04. This dataset includes user ratings for video games,
which provides a suitable test for item recommendations in the entertainment domain.

The Amazon Bundle dataset contains 14,240 items and 2,376 sessions across three subcategories: Electronics, Clothing, and Food. The average session length is 6.73, with a density indicator of 1.12. This dataset includes session data with
explicitly annotated user intents, making it particularly valuable for testing intent-based recommendation models.

For all datasets, we preprocessed the data by ordering the user interactions chronologically and dividing them into sessions.
For ML-1M and Amazon Games, the interactions are grouped into sessions based on daily interactions. The Amazon Bundle dataset is already sessionized and annotated, so it is used directly. Each dataset is split into training, validation,
and test sets using a chronological approach. Specifically, the first 80\% of sessions are used for training, the subsequent 10\% for validation, and the final 10\% for testing. This split ensures that the model is trained on earlier data and tested on more recent interactions, mimicking real-world recommendation systems.

\subsection{Baselines}
To evaluate the performance of DUIP, we compare it with 11 baseline models categorized into three types: conventional methods, deep learning-based methods, and LLM-based methods. These baselines represent a wide spectrum of approaches from traditional techniques to advanced neural network-based models, each capturing user intent and session information in different ways. Below is a detailed list of these baselines:

- Mostpop: Recommends the most popular items based on user interactions.

- SKNN \cite{jannach2017recurrent}: Recommends session-level similar items using session-based nearest neighbor search.

- FPMC \cite{rendle2010factorizing}: A matrix factorization method that incorporates the first-order Markov chain to model user-item interactions.

- NARM \cite{li2017neural}: An RNN-based model with attention mechanisms to capture the main user intent from hidden states.

- STAMP \cite{liu2018stamp}: Learns the user’s primary intent by focusing on the impact of the last item in the context.

- GCE-GNN \cite{wang2020global}: Uses both local and global graphs to learn item representations and identify the main intent of a session.

- MCPRN \cite{wang2019modeling}: Models users' multiple purposes to derive a final session representation.

- HIDE \cite{li2022enhancing}: Splits item embeddings into multiple chunks, each representing a specific intention to learn diverse user intents.

- Atten-Mixer \cite{zhang2023efficiently}: Learns multi-granularity consecutive user intents for more accurate session representations.

- UniSRec \cite{hou2022towards}: A cross-domain model that uses item descriptions to learn transferable representations across different domains.

- NIR \cite{wang2023zero}: Adopts zero-shot prompting to recommend the next item.

\subsection{Results and Analysis}

\begin{table}[h!]
\begin{minipage}[t]{0.48\textwidth}
	\caption{Performance comparison on ML-1M}
        \label{tab:table1}
	\centering	
	\setlength{\tabcolsep}{2pt} 
	\begin{tabular}{ccccc}
		\toprule
		\textbf{Model} & \textbf{HR@1} & \textbf{HR@5} & \textbf{NDCG@1} & \textbf{NDCG@5} \\
		\midrule
		MostPop & 0.0004 & 0.0070 & 0.0004 & 0.0053 \\
		
		SKNN & 0.1270 & 0.3600 & 0.1270 & 0.2530 \\
		
		FPMC & 0.1132 & 0.3748 & 0.1132 & 0.2464 \\
		
		NARM & 0.1692 & 0.5230 & 0.1692 & 0.3501 \\
		
		STAMP & 0.1584 & 0.5078 & 0.1584 & 0.3367 \\
		
		GCE-GNN & 0.1312 & 0.4748 & 0.1312 & 0.3044 \\
		
		MCPRN & 0.1434 & 0.4788 & 0.1434 & 0.3157 \\
		
		HIDE & 0.1498 & 0.4998 & 0.1498 & 0.3256 \\
		
		Atten-Mixer & 0.1490 & 0.4932 & 0.1490 & 0.3216 \\
		
		UniSRec & 0.0508 & 0.2508 & 0.0508 & 0.1459 \\
		
		NIR & 0.0572 & 0.2326 & 0.0572 & 0.1436 \\
		
		\textbf{DUIP} & \textbf{0.1883} & \textbf{0.5348} & \textbf{0.1883} & \textbf{0.3674} \\
		\bottomrule
	\end{tabular} \hspace{.3cm}
     \end{minipage}%
     \hfill
     \begin{minipage}[t]{0.48\textwidth}
	\caption{Performance comparison on Games}
        \label{tab:table2}
	\centering	
	\setlength{\tabcolsep}{2pt} 
	\begin{tabular}{ccccc}
		\toprule
		\textbf{Model} & \textbf{HR@1} & \textbf{HR@5} & \textbf{NDCG@1} & \textbf{NDCG@5} \\
		\midrule
		SKNN & 0.0020  & 0.0020  & 0.0020  & 0.0020  \\
		
		FPMC & 0.0498  & 0.2564  & 0.0498  & 0.1508  \\
		
		NARM & 0.0572  & 0.2574  & 0.0572  & 0.1534  \\
		
		STAMP & 0.0556  & 0.2586  & 0.0556  & 0.1555  \\
		
		GCE-GNN & 0.0692  & 0.2744  & 0.0692  & 0.1701  \\
		
		MCPRN & 0.0522  & 0.2416  & 0.0522  & 0.1432  \\
		
		HIDE & 0.0696  & 0.2694  & 0.0696  & 0.1662  \\
		
		Atten-Mixer & 0.0530  & 0.2472  & 0.0530  & 0.1475  \\
		
		UniSRec & 0.0544  & 0.2512  & 0.0544  & 0.1482  \\
		
		NIR & 0.1168 & 0.3406 & 0.1168 & 0.2310 \\
		
		\textbf{DUIP} & \textbf{0.1732}  & \textbf{0.3729}  & \textbf{0.1732}  & \textbf{0.2561}  \\
		\bottomrule
	\end{tabular}
      \end{minipage}%
\end{table}
The experimental results demonstrate that DUIP significantly outperforms a range of baseline models across multiple datasets, highlighting its effectiveness in addressing the dynamic nature of user preferences. The combination of LSTM-based dynamic user intent modeling and LLM-based next-item prediction allows DUIP to capture and adapt to evolving user behaviors, which is a major advantage over traditional methods and even some deep learning-based models.

As shown in Table \ref{tab:table1}, on the ML-1M dataset, DUIP delivered impressive performance, especially in HR@1 and HR@5, showing a clear improvement over models such as SKNN, NARM, and STAMP, which rely on static representations of user preferences or sequential patterns without the ability to adapt dynamically to shifts in intent. The LSTM component of DUIP processes user-item interaction sequences, capturing long-term dependencies in user behavior, while the LLM provides a powerful mechanism for generating context-aware predictions. This synergy enables DUIP to not only recommend relevant items but also rank them appropriately, as evidenced by its superior NDCG scores.


As shown in Table \ref{tab:table2}, the Games dataset, which presents a challenge due to sparse data and the variety of user interests, further showcases DUIP’s strength. Here, DUIP outperformed the baselines in both HR@1 and NDCG@5, with a notable improvement in NDCG@1, indicating its ability to provide both relevant and well-ranked recommendations. The LSTM’s ability to continuously update the user’s intent based on recent interactions gives DUIP an edge in adapting to the rapidly changing interests of users, especially in domains like entertainment, where preferences can vary significantly over short periods of time.

\begin{table}[h!]
	\caption{Performance comparison on Bundle}
        \label{tab:table3}
	\centering	
	\setlength{\tabcolsep}{4pt} 
	\begin{tabular}{ccccc}
		\toprule
		\textbf{Model} & \textbf{HR@1} & \textbf{HR@5} & \textbf{NDCG@1} & \textbf{NDCG@5} \\
		\midrule
		MostPop & – & 0.0042  & – & 0.0021  \\
		
		FPMC & 0.0398  & 0.2475  & 0.0398  & 0.1395  \\
		
		NARM & 0.0322  & 0.2322  & 0.0322  & 0.1303  \\
		
		STAMP & 0.0365  & 0.2352  & 0.0365  & 0.1339  \\
		
		GCE-GNN & 0.0360  & 0.2237  & 0.0360  & 0.1267  \\
		
		MCPRN & 0.0360  & 0.2352  & 0.0360  & 0.1490  \\
		
		HIDE & 0.0458  & 0.2585  & 0.0458  & 0.1495  \\
		
		Atten-Mixer & 0.0525  & 0.2644  & 0.0525  & 0.1549  \\
		
		UniSRec & 0.0496  & 0.2402  & 0.0496  & 0.1430  \\
		
		NIR & 0.0975 & 0.2832 & 0.0975 & 0.1939 \\
		
		\textbf{DUIP} & \textbf{0.1233}  & \textbf{0.3001}  & \textbf{0.1233}  & \textbf{0.2217}  \\
		\bottomrule
	\end{tabular}
\end{table}

As shown in Table \ref{tab:table3}, The performance on the Bundle dataset, which is more session-based and has annotated user intents, also reflects DUIP’s capability in handling session-specific dynamics. Although the performance gap was narrower compared to ML-1M and Games, DUIP still outperformed the baselines in terms of HR@1 and HR@5, demonstrating that the model can effectively work with session-level data and adapt its recommendations to the immediate needs of users, even in sparse interaction environments.

\section{CONCLUSION}
In this paper, we proposed DUIP, a novel recommendation framework that integrates LSTM-based dynamic user intent modeling with LLM-based item prediction. The results of our extensive experiments across three diverse datasets—ML-1M, Games, and Bundle—demonstrate that DUIP significantly outperforms existing baseline models, including traditional methods, deep learning-based approaches, and LLM-based models. By capturing the dynamic nature of user intent and leveraging LSTM to model sequential interactions, DUIP is able to generate highly personalized, context-aware recommendations. The integration of LLMs further enhances the model’s ability to generate accurate and relevant predictions, ensuring that recommendations are not only personalized but also reflect real-time shifts in user preferences.

The superior performance of DUIP across HR@1, HR@5, NDCG@1, and NDCG@5 metrics highlights its ability to improve both top-k recommendation accuracy and ranking quality, which are crucial in modern recommendation systems. This makes DUIP a promising solution to long-standing challenges such as cold-start and dynamic user intent prediction. Despite its impressive results, further research can focus on optimizing the model for handling sparse datasets and improving computational efficiency, especially for large-scale real-world applications.

Looking ahead, there are several potential directions for future research and improvement of DUIP. First, integrating additional multi-modal data (such as images, text, and social context) could further enhance the model’s adaptability and understanding of user intent. Cross-domain recommendations could also be explored by expanding the model’s capability to transfer knowledge across different types of recommendation tasks, such as between movies, products, and music. Furthermore, addressing the scalability of the model in handling large datasets and real-time user interactions could make DUIP even more effective in production environments. Lastly, enhancing the real-time adaptation mechanism of DUIP through online learning techniques could allow it to update its recommendations instantaneously as new user data becomes available.

\bibliography{report} 
\bibliographystyle{spiebib} 

\end{document}